\documentclass[showpacs,aps,amsmath,amssymb,twocolumn]{revtex4}
\usepackage[latin1]{inputenc}
\usepackage{graphicx}
\usepackage{bm}
\usepackage{amsmath}
\usepackage{array}
\usepackage{braket}
\usepackage{microtype}

\renewcommand{\baselinestretch}{1.0}
\begin{document}

\renewcommand{\baselinestretch}{1.0}

\title{Unified quantum density matrix description of coherence and polarization}

\author{Bert\'{u}lio de Lima Bernardo$^{1,2}$}

\affiliation{$^{1}$Departamento de F\'{i}sica, Universidade Federal de Pernambuco, 50670-901 Recife-PE, Brazil\\
$^{2}$Departamento de F\'{i}sica, Universidade Federal de Campina Grande, Caixa Postal 10071, 58109-970 Campina Grande-PB, Brazil}

\email{bertulio.fisica@gmail.com}

\begin{abstract}

The properties of coherence and polarization of light has been the subject of intense investigations and form the basis of many technological applications. These concepts which historically have been treated independently can now be formulated under a single classical theory. Here, we derive a quantum counterpart for this theory, with basis on a density matrix formulation, which describes jointly the coherence and polarization properties of an ensemble of photons. The method is used to show how the degree of polarization of a specific class of mixed states changes on propagation in free space, and how an interacting environment can suppress the coherence and polarization degrees of a general state. This last application can be particularly useful in the analysis of decoherence effects in optical quantum information implementations.

\end{abstract}

\maketitle


\section{Introduction}

Coherence and polarization are undoubtedly two of the most important properties of light. In general terms, the coherence of an optical field can be understood as the ability to produce interference, as remarkably demonstrated by Young in his famous double-slit experiment, and theoretically developed by the works of Fresnel in the context of waves \cite{hecht}. Another important development in the coherence theory was the one made by Glauber and Sudarshan, which established the connections about the coherence properties of light with the concept of photon statistics in a quantum mechanical scenario \cite{glauber,glauber2,sud}. Conversely, the modern study of the polarization properties was introduced by Stokes, who proposed a set of parameters to completely describe the polarization state of a random electromagnetic wave; the so-called Stokes parameters \cite{mc,mc2}, that can also be extended to the quantum realm \cite{agrawal}. Together, these two concepts form the basis of numerous applications of light in microscopy \cite{moerner}, criptography \cite{bennet,ekert}, metrology \cite{katori}, astronomy \cite{hanbury,abbott}, as well as in future quantum information technologies \cite{nielsen,kok}.

Although the importance of the theories of coherence and polarization, their theoretical descriptions have historically been developed independently \cite{born,brosseau,mandel}. However, since the last decade the study of these two apparently distinct properties could be established into a single formulation through the unified theory of coherence and polarization introduced by Wolf \cite{wolf}. In this seminal work, it was shown that both coherence and polarization of a random electromagnetic beam could be understood as manifestations of the correlations between fluctuations of the optical field. In this respect, coherence manifests itself from correlations between fluctuations of the electric field of a light beam at two or more points in space, whereas polarization arises from the correlations of the optical field components at a single point in space \cite{wolf2}.  

Since the publication of the unified theory, many other advances have been made towards a complete understanding of this problem. For example, the introduction of the generalized Stokes parameters \cite{koro}, the description of the polarization change of partially coherent electromagnetic beam upon propagation in free space \cite{koro2,salem}, and in the turbulent atmosphere \cite{koro3,roy}, just to mention a few. Nevertheless, almost all these works have been limited to the scope of the classical electromagnetic theory \cite{lind,lepp}. In fact, there have been some recent works extending the classical unification theory to the realm of quantum mechanics by direct quantization of the electromagnetic field \cite{lahiri,lahiri2}. So far, this extension did not provide a significant clarification of the problem, when compared to the classical counterpart, maybe because the state of the field is characterized in the Fock space, which sometimes makes the physical intuition less precise and, depending on the environment in which the system is inserted, it is difficult to write an appropriate Hamiltonian to account for the time evolution of the system \cite{agrawal2}.  

In this work, we derive a unified quantum mechanical description of coherence and polarization from first principles, that is to say, without direct reference to the classical theory. As we shall see, the central element in this formalism is the density matrix of the system written directly in terms of the position and polarization Hilbert spaces. This last point is the responsible for making the method relatively simple when describing the behavior of a general ensemble of photons on propagation in free space, as well as under the action of an interacting environment. Indeed, we provide some applications of the model to demonstrate how a partially coherent ensemble of photons change the degree of polarization when propagating in free space, and how decoherence and depolarization take place when photons are subjected to an environment whose constituents can be refractive and birefringent. Since all these examples are presented by means of simple quantum-mechanical arguments, the present description can be particularly valuable in the study of environmental disturbance in optical quantum information processes, in which the properties of coherence and polarization play a fundamental role.

\section{Theory}

To start with, we derive an expression for the degree of spatial coherence of light in a context similar to the one used to derive the classical theory \cite{wolf}. In doing so, let us consider a Young's double-slit experiment which consists in an ensemble of photons propagating close to the z-axis which are mostly blocked by a mask with two small openings on it. After this stage, the positions of the photons that passed through the slits are permanently registered by a distant detection screen, as shown in Fig. 1. Let $\ket{0}$ and $\ket{1}$ denote the quantum states of the photons which passed through the slits $Q_{0}$ and $Q_{1}$, respectively, and $\ket{H}$ and $\ket{V}$ the states of the photons linearly polarized along the horizontal (x-axis) and vertical (y-axis) directions, respectively. In this scenario, we can write the general quantum state of the photons in the form
\begin{equation}
\label{1}
\ket{\psi} = a\ket{H,0} + b\ket{H,1} + c\ket{V,0} + d\ket{V,1},
\end{equation}
with $|a|^{2} + |b|^{2} + |c|^{2} + |d|^{2} = 1$,
in order to specify simultaneously both the slit in which the photon passes through and the state of polarization. Also, we can write the density matrix for this system as $\hat{\rho}=\ket{\psi}\bra{\psi}$, which provides a $4 \times 4$ matrix in the following format:  
\begin{equation}
\label{2}
\hat{\rho} =
\begin{pmatrix}
\rho_{11} & \rho_{12} & \rho_{13} & \rho_{14} \\
\rho_{21} & \rho_{22} & \rho_{23} & \rho_{24} \\
\rho_{31} & \rho_{32} & \rho_{33} & \rho_{34} \\
\rho_{41} & \rho_{42} & \rho_{43} & \rho_{44}
  \end{pmatrix}
 =
\begin{pmatrix}
|a|^{2} & ab^{*} & ac^{*} & ad^{*} \\
ba^{*} & |b|^{2} & bc^{*} & bd^{*} \\
ca^{*} & cb^{*} & |c|^{2} & cd^{*} \\
da^{*} & db^{*} & dc^{*} & |d|^{2}
\end{pmatrix},
\end{equation}
where the asterisk denotes complex conjugation.
\begin{figure}[htb]
\begin{center}
\includegraphics[height=1.45in]{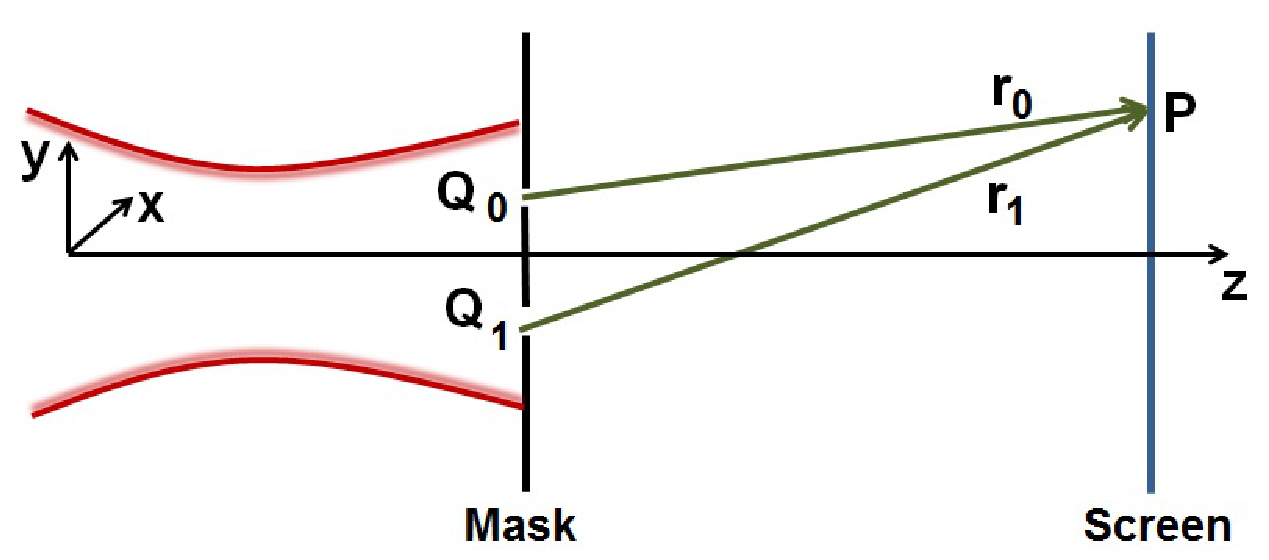}
\end{center}
\label{F1}
\caption{(Color online) Scheme of the double-slit experiment. An ensemble of photons impinges a mask containing two slits, $Q_{0}$ and $Q_{1}$, rendering two possible paths to each of them, which are afterwards detected on the screen.}    
\end{figure}

Now, if we are interested in computing the probability density $\rho(P)$ to find a photon at a point $P$ on the detection screen, keeping in mind that horizontally polarized photons do not interfere with vertically polarized ones, we have that
\begin{equation}
\label{3}
\rho(P)= \braket{H,P|\rho|H,P} + \braket{V,P|\rho|V,P},
\end{equation}
where $\ket{H,P}$ and $\ket{V,P}$ represent the states of photons localized at $P$ with horizontal and vertical polarizations, respectively. Assuming that the size of the slits is much smaller than the wavelength of the photons, we can consider that after passing through a given slit the wavefunction of the photons are spherical waves. Therefore, the probability amplitudes of finding a photon at $P$ with horizontal (vertical) polarization which passed through the slit $Q_{0}$ ($Q_{1}$) are given, respectively, by
\begin{equation}
\label{4}
\psi^{(P)}_{H,V}(r_{0}) = \braket{H,P|H,0} = \braket{V,P|V,0} = \frac{e^{ikr_{0}}}{r_{0}}
\end{equation}
and
\begin{equation}
\label{5}
\psi^{(P)}_{H,V}(r_{1}) = \braket{H,P|H,1} = \braket{V,P|V,1} = \frac{e^{ikr_{1}}}{r_{1}},
\end{equation}
with $i$ and $k$ being the imaginary unity and the wavenumber, respectively. The parameters $r_{0}$ and $r_{1}$ are the distances from the slits $Q_{0}$ and $Q_{1}$ to the point $P$, respectively. By substitution of Eqs.~\eqref{4} and~\eqref{5} into Eq.~\eqref{3}, and cancelling out terms with inner products between horizontal and vertical polarization states, we obtain that
\begin{eqnarray}
\label{6}
\rho(P)&=&\frac{\rho_{11} + \rho_{33}}{r^{2}_{0}} + \frac{\rho_{22} + \rho_{44}}{r^{2}_{1}} \nonumber\\
 &+&  \frac{2 Re[(\rho_{21} + \rho_{34}) e^{ik(r_{0}-r_{1})}]}{r_{0}r_{1}} ,
\end{eqnarray}
where we used the fact that $\rho$ is Hermitian, $\rho_{mn} = \rho^{*}_{nm}$, and $Re$ denotes the real part.

Let us visualize Eq.~\eqref{6} under a different perspective. Observe that if the slit $Q_{1}$ is closed, the amplitudes $b$ and $d$ are null in Eq.~\eqref{1}, therefore, Eq.~\eqref{6} reduces to
\begin{equation}
\label{7}
\rho_{0}(P)=\frac{\rho_{11} + \rho_{33}}{r^{2}_{0}},
\end{equation}
which represents the probability density of finding a photon that emerged exclusively from $Q_{0}$ at $P$. Similarly, the probability density of finding a photon that emerged from $Q_{1}$ at $P$ is given by
\begin{equation}
\label{8}
\rho_{1}(P)=\frac{\rho_{22} + \rho_{44}}{r^{2}_{1}}.
\end{equation}
In this context, Eq.~\eqref{6} can be rewritten as
\begin{equation}
\label{9}
\rho(P)= \rho_{0}(P) + \rho_{1}(P) + 2 \sqrt{\rho_{0}(P)} \sqrt{\rho_{1}(P)} Re[\mu e^{ik(r_{0}-r_{1})}],
\end{equation}
where the parameter $\mu$ is given by
\begin{equation}
\label{10}
\mu = \frac{\rho_{12} + \rho_{34}}{\sqrt{\rho_{11} + \rho_{33}} \sqrt{\rho_{22} + \rho_{44}}}.
\end{equation}
The first two terms in Eq.~\eqref{9} correspond to the sum of the individual probability densities of the photons which passed through each slit, and the last term is responsible for the interference pattern on the detection screen. Note that the parameter that dictates the prominence of the interference pattern in this system is $\mu$, which we define as the {\it degree of coherence}. Therefore, this parameter can be measured by detecting the patterns due to the photons which emerge from each slit separately, and the pattern formed when both slits are open, by means of Eq.~\eqref{9}.  

By substitution of the amplitudes $a$, $b$, $c$ and $d$ in Eq.~\eqref{10}, and using the Cauchy-Schwarz inequality, it is easy to show that $0\leq|\mu|\leq1$. Since the interference term is maximum when $|\mu| = 1$, we say that the ensemble of photons is {\it completely coherent} with respect to the slits $Q_{0}$ and $Q_{1}$. On the other hand, for $\mu =0$, the interference pattern is completely destroyed and we say that the ensemble of photons is {\it completely incoherent} with respect to the slits. In the intermediate cases, $0< |\mu| < 1$, we say that the photons are {\it partially coherent}. Despite the fact that we used a purely quantum mechanical method to derive the expression for the degree of coherence, Eq.~\eqref{10}, it has an interesting mathematical similarity with the one found in the classical theory (See Eq. (8) in Ref. \cite{wolf}).

Now we turn to the analysis of the polarization in this system, and to the derivation of an expression for the degree of polarization. Initially, let us concentrate only on the photons emerging from the slit $Q_{0}$. At this point, we can define the quantum version of the Stokes parameters as given by (see a similar analysis in Ref. \cite{alte})  
\begin{equation}
\label{11}
S^{(0)}_{0} = tr[\ket{H,0}\bra{H,0}\rho] + tr[\ket{V,0}\bra{V,0}\rho]=\rho_{11}+\rho_{33},
\end{equation}  
\begin{equation}
\label{12}
S^{(0)}_{1} = tr[\ket{H,0}\bra{H,0}\rho] - tr[\ket{V,0}\bra{V,0}\rho]=\rho_{11}-\rho_{33},
\end{equation}
\begin{equation}
\label{13}
S^{(0)}_{2} = tr[\ket{H,0}\bra{V,0}\rho] + tr[\ket{V,0}\bra{H,0}\rho]=\rho_{13}+\rho_{31},
\end{equation}
\begin{equation}
\label{14}
S^{(0)}_{3} = i\{tr[\ket{V,0}\bra{H,0}\rho] - tr[\ket{H,0}\bra{V,0}\rho]\}=i (\rho_{13}-\rho_{31}),
\end{equation}
where $tr$ denotes the trace. Observe that, contrary to the second quantization formalism, in which the Stokes parameters are operators \cite{agrawal,luis,luis2}, here they appear simply as numbers playing a role similar to that of the classical electromagnetic theory. Physically, the Stokes parameters above represent the ensemble average of the identity operator and the three Pauli operators in the basis $\{\ket{H},\ket{V}\}$ \cite{brosseau1}, but only for the photons which emerge from slit $Q_{0}$. In this form, they can be measured with an appropriate combination of linear polarizers and a $\pi/2$ phase shifter (See Refs. \cite{schaef} and \cite{alte} for a classical and quantum approach to this problem, respectively).

Accordingly, the degree of polarization at $Q_{0}$ can be defined, and measured, in agreement with the following relation \cite{brosseau2,bjork}:
\begin{equation}
\label{15}
p_{0}= \frac{\sqrt{(S^{(0)}_{1})^{2}+(S^{(0)}_{2})^{2}+(S^{(0)}_{3})^{2}}}{S^{(0)}_{0}},
\end{equation}
which is a real quantity and its range is $0 \leq p_{0} \leq 1$ \cite{wolf2}. After substitution of Eqs.~\eqref{11},~\eqref{12},~\eqref{13} and ~\eqref{14} into Eq.~\eqref{15}, and some algebra, it can be written as
\begin{equation}
\label{16}
p_{0}= \sqrt{1-\frac{4(\rho_{11}\rho_{33}-\rho_{13}\rho_{31})}{(\rho_{11}+\rho_{33})^{2}}},
\end{equation}
which is our final expression for the degree of polarization. With this definition, we have that: (i) if $0<p_{0}<1$, the ensemble is said to be {\it partially polarized}; (ii) if $p_{0}=0$, the ensemble is {\it unpolarized}, and (iii) if $p_{0}=1$, the ensemble is totally {\it polarized}.  In a similar fashion, one can show that the degree of polarization of the photons which pass through the slit $Q_{1}$ is given by
\begin{equation}
\label{17}
p_{1}= \sqrt{1-\frac{4(\rho_{22}\rho_{44}-\rho_{24}\rho_{42})}{(\rho_{22}+\rho_{44})^{2}}}.
\end{equation}
Again, we call attention to the mathematical similarity between the expression of the degree of polarization derived here, with basis only on a quantum mechanical background, and the one found by means of classical methods \cite{wolf}.

So far, we have discussed the properties of coherence and polarization based on the pure state of Eq.~\eqref{1}. However, for a general mixed state
\begin{equation}
\label{18}
\hat{\rho} = \sum_{i}w_{i} \rho^{(i)} =\sum_{i}w_{i}\ket{\psi^{(i)}}\bra{\psi^{(i)}},
\end{equation}
with $w_{i}$ as the fractional populations of each pure state $\rho^{(i)}$ contained in the ensemble of photons, we can write the density matrix for the system as
\begin{equation}
\label{19}
\hat{\rho} =
\begin{pmatrix}
\rho_{11} & \rho_{12} & \rho_{13} & \rho_{14} \\
\rho_{21} & \rho_{22} & \rho_{23} & \rho_{24} \\
\rho_{31} & \rho_{32} & \rho_{33} & \rho_{34} \\
\rho_{41} & \rho_{42} & \rho_{43} & \rho_{44}
  \end{pmatrix},
\end{equation}
with each element given by $\rho_{nm} = \sum_{i} w_{i} \rho^{(i)}_{nm}$. In this form, all the derivations developed above are equally valid for a mixed density matrix, including the expressions for the degree of coherence and polarization of Eqs.~\eqref{10},~\eqref{16} and~\eqref{17}.    

Now we present some applications of the equations derived above in order to elucidate their physical meaning. Consider, for example, a pure ensemble of horizontally polarized photons with equivalent probabilities of passing through both slits. This is described by the state $\ket{\psi} = \frac{1}{\sqrt{2}}(\ket{H,0} + \ket{H,1}$, such that the degrees of coherence as well as polarization are found to reach their maximum value, namely, $\mu=1$, $p_{0}=1$ and $p_{1}=1$. These are expected results since we have no information about which slit the photons passed, and all of them have a well defined polarization. In fact, for a general pure state described by Eq.~\eqref{1}, the two aspects that determine the degree of coherence are: (i) the relation between the probabilities of a photon to emerge from $Q_{0}$ and $Q_{1}$, and (ii) the similarity between horizontally and vertically polarized photons with respect to the phase relation of the passage through the two slits. In the first case, the more distributed the probabilities of the photon to pass through each slit, the larger the degree of coherence. In the second case, the closer the relative phases between the passage through $Q_{0}$ and $Q_{1}$ for photons with polarization horizontal and vertical, the larger the degree of coherence. In regards to the degrees of polarization, it is easy to see that they reach their maximum for a pure state, $p_{0} = p_{1} = 1$. This is also expected since, independent of the amplitudes $a$, $b$, $c$ and $d$, we always have a well defined polarization in the H-V basis for the state $\ket{\psi}$ in Eq.~\eqref{1}.

Note that the present discussion is also valid for states whose coherent properties are entangled with the polarization properties. For instance, the state $\ket{\psi} = a\ket{H,0} + d\ket{V,1}$ provides $\mu=0$, $p_{0}=1$ and $p_{1}=1$. The reason why the degree of coherence is null is because the polarization gives information about the path of the photons, eliminating the interference pattern. The degree of polarization is maximum because the polarization is completely defined at each opening, namely, horizontal at $Q_{0}$ and vertical at $Q_{1}$.  

The situation is much richer for mixed states. Here, we want to analyze two particular examples. First, consider the mixed state
\begin{eqnarray}
\label{20}
\hat{\rho} &=& 1/4[\ket{H,0}\bra{H,0}+\ket{V,0}\bra{V,0}\\ \nonumber
&+&\ket{H,1}\bra{H,1}+\ket{V,1}\bra{V,1}],
\end{eqnarray}
which represents a completely random ensemble of photons with equal probability of being horizontally and vertically polarized, and passing through the slits $Q_{0}$ and $Q_{1}$. Then, for obvious reasons, the degrees of coherence and polarization are null: $\mu=0$, $p_{0}=0$ and $p_{1}=0$. The next example is particularly interesting. It represents a mixed state whose coherence and polarization degrees of freedom are separable:
\begin{eqnarray}
\label{21}
\hat{\rho}&=& \hat{\rho}_{p} \otimes \hat{\rho}_{c} \\ \nonumber
&=&  1/2[\ket{H}\bra{H}+\ket{V}\bra{V}] \\ \nonumber
&\otimes& 1/2[\ket{0}\bra{0}+\ket{1}\bra{1} + \ket{0}\bra{1}+\ket{1}\bra{0} ].      
\end{eqnarray}
The degrees of coherence and polarization for this state satisfy $\mu=1$, $p_{0}=0$ and $p_{1}=0$. That is, the ensemble of photons is completely unpolarized at each slit; however, it is completely coherent with respect to them. Wolf called attention to this unusual behavior when the classical theory was developed \cite{wolf}.

\section{Polarization change on propagation in free space}

In this section, we use the theory that we developed to show how the degree of polarization of a mixed ensemble of photons is changed upon propagation along the $z$ direction. Let us consider the ensemble initially prepared (at $z=0$) in the following mixed state:
\begin{eqnarray}
\label{22}
\hat{\rho}(0) &=& w_{1}(0)\hat{\rho}_{1} + w_{2}(0)\hat{\rho}_{2}
\\ \nonumber
&=& w_{1}(0)\ket{\psi_{1}}\bra{\psi_{1}} + w_{2}(0)\ket{\psi_{2}}\bra{\psi_{2}},
\end{eqnarray}
with $\ket{\psi_{1}} = \frac{1}{\sqrt{2}}(\ket{H,0}+\ket{H,1})$ and $\ket{\psi_{2}} = \frac{1}{\sqrt{2}}(\ket{V,0}+\ket{V,1})$, which is composed of two groups of pure states that we assume to be, in the beginning, equally probable to be detected at the points $Q_{0}$ and $Q_{1}$. Namely, $w_{1}(0) = w_{2}(0) = 1/2$ \cite{note}. The first group is horizontally polarized, with equal probability of detecting a photon at $Q_{0}$ and $Q_{1}$. The second is vertically polarized, but also with equal probability of detecting a photon at the two referred points. Moreover, let us assume that the spatial wavefunctions of each group perpendicular to the direction of propagation, $y$, are Gaussians with different widths, which spread on propagation due to diffraction (See Fig. 2):
\begin{equation}
\label{23}
\psi_{1}(y,z) =\left( \frac{1}{\sqrt{2 \pi} \sigma_{1}(z)} \right)^{\frac{1}{2}} \exp \left(- \frac{y^{2}}{4 \sigma_{1}^{2}(z)} \right),
\end{equation}
and
\begin{equation}
\label{24}
\psi_{2}(y,z) =\left( \frac{1}{\sqrt{2 \pi} \sigma_{2}(z)} \right)^{\frac{1}{2}} \exp \left(- \frac{y^{2}}{4 \sigma_{2}^{2}(z)} \right),
\end{equation}
where $\sigma_{1}(z)$ and $\sigma_{2}(z)$ are the respective widths of the wavefunctions. Note that, for a given point along the $y$ direction the relative probabilities to find a photon described by $\psi_{1}$ (with horizontal polarization) or $\psi_{2}$ (with vertical polarization) change on propagation. This, as we shall see, is the physical reason for which the degree of polarization varies with $z$.

The $z$-dependent density matrix of this system in the basis $\{\ket{H,0};\ket{H,1};\ket{V,0};\ket{V,1}\}$ is given by:
\begin{equation}
\label{25}
\hat{\rho} (z)= \frac{1}{2}
\begin{pmatrix}
w_{1}(z) & w_{1}(z) & 0 & 0 \\
w_{1}(z) & w_{1}(z) & 0 & 0 \\
0 & 0 & w_{2}(z) & w_{2}(z) \\
0 & 0 & w_{2}(z) & w_{2}(z)
  \end{pmatrix}.
\end{equation}
Thus, the degrees of polarization at the points $Q_{0}$ and $Q_{1}$ are given by:
\begin{equation}
\label{26}
p_{0}=p_{1} = p = \sqrt{1-\frac{4 w_{1}(z) w_{2}(z)}{[w_{1}(z)+w_{2}(z)]^{2}}}.
\end{equation}
At this point we call attention to the fact that the fractional populations $w_{1}(0)$ and $w_{2}(0)$ to find each subensemble at $Q_{0}$ and $Q_{1}$, at $z=0$, introduced in Eq.~\eqref{22} are numbers which are determined at the moment of the creation of the ensemble. Then, it should be clear that the dependence on $z$ shown in Eq.~\eqref{25} reflects that the probabilities to obtain a photon from each group changes upon propagation because of the independent evolution of $\psi_{1}(y,z)$ and $\psi_{2}(y,z)$.  
\begin{figure}[htb]
\begin{center}
\includegraphics[height=1.5in]{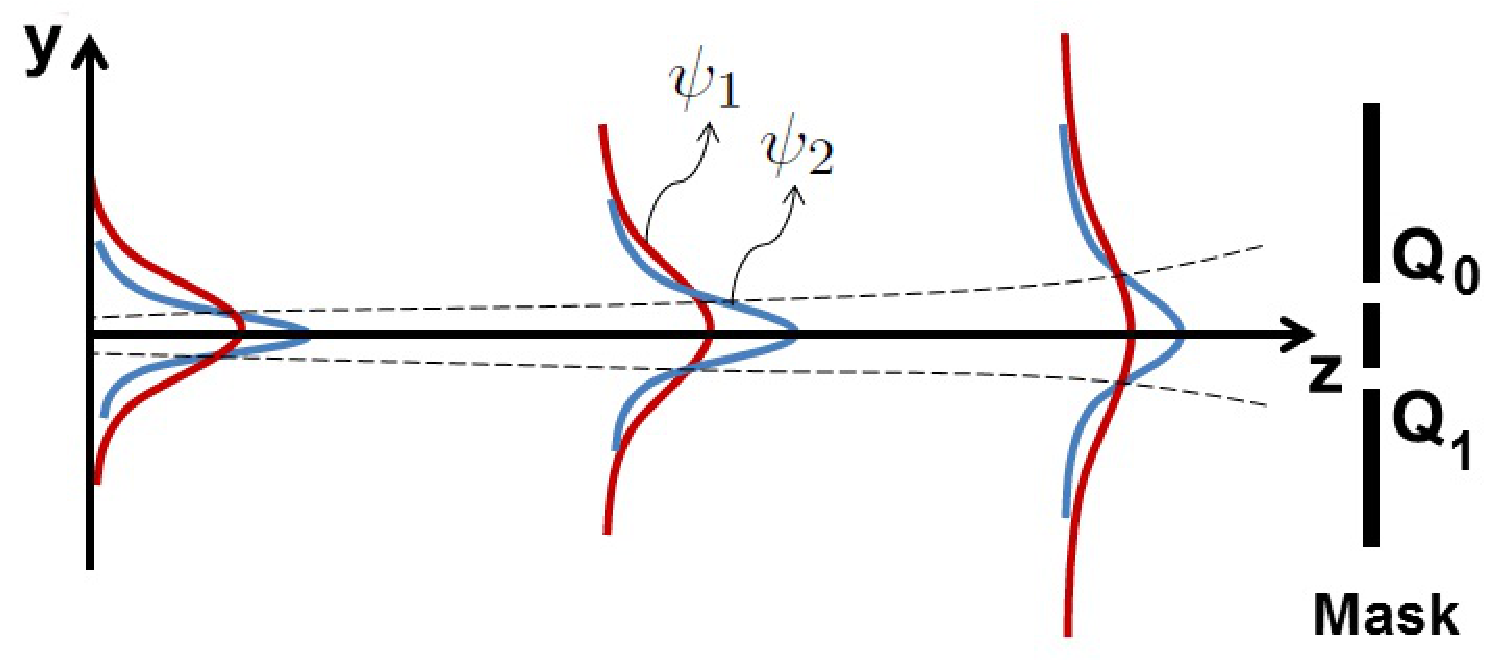}
\end{center}
\label{F1}
\caption{(Color online) Propagation of an ensemble of photons composed of two subensembles with orthogonal linear polarization states. The spatial wavefunctions of both have a Gaussian profile with different widths, which evolve independently. The dashed lines indicate the points in which both subensembles are equally probable to be detected.}    
\end{figure}

In order to estimate the probabilities to obtain each group of photons, we will assume that the subensembles comprise two Gaussian beams whose intensity profiles (probability density) are given by \cite{mandel,siegman}:
\begin{equation}
\label{27}
W_{1}(y,z) =\left( \frac{\sigma_{1}(0)} {\sigma_{1}(z)}\right)^{2} \exp \left(- \frac{2 y^{2}}{\sigma^{2}_{1}(z)} \right),
\end{equation}
and
\begin{equation}
\label{28}
W_{2}(y,z) =\left( \frac{\sigma_{2}(0)} {\sigma_{2}(z)}\right)^{2} \exp \left(- \frac{2 y^{2}}{\sigma^{2}_{2}(z)} \right),
\end{equation}
where $\sigma_{1}(0)$ and $\sigma_{2}(0)$ are the minimum widths of  $\psi_{1}(r,z)$ and $\psi_{2}(r,z)$, respectively, which we assume to be at $z=0$. We also have that $\sigma_{j}(z)=\sigma_{j}(0) \sqrt{1+(z/z_{j})^{2}}$, with $j=1,2$. The parameters $z_{j}$ are the Rayleigh lengths of the beams described by $\psi_{1}(r,z)$ and $\psi_{2}(r,z)$.

For the sake of simplicity, let us assume that the distance between the points $Q_{0}$ and $Q_{1}$ and the $z$-axis are much smaller than the minimum width of the beams, such that we can write  
\begin{equation}
\label{29}
W_{j}(z) \approx \left( \frac{\sigma_{j}(0)} {\sigma_{j}(z)}\right)^{2} = \left[ 1+ \left( \frac{z}{z_{j}}\right)^{2} \right]^{-1},
\end{equation}
which after normalization we find that the matrix elements in Eq.~\eqref{25} are given by
\begin{equation}
\label{30}
w_{j}(z)  = \frac{\left[ 1+ \left( \frac{z}{z_{j}}\right)^{2} \right]^{-1}} {\sum_{j=1,2} \left[ 1+ \left( \frac{z}{z_{j}}\right)^{2} \right]^{-1}}.
\end{equation}
Observe that $w_{j}(0) = 1/2$, which are the fractional populations that we defined at the ensemble creation. Finally, if we substitute Eq.~\eqref{30} into Eq.~\eqref{26}, we can find how the degree of polarization changes as the photons propagate along the $z$-axis. Fig. 3 shows the degree of polarization of Eq.~\eqref{26}, with $w_{j}(z)$ as given in Eq.~\eqref{30}, as a function of the propagation direction $z$, for $z_{2}=2 z_{1}$. The behavior is similar to the one found by using the classical theory applied to the Gaussian Schell-model \cite{james}. In this regard, it is important to emphasize that despite the similarity between the results obtained from the classical formalism and the quantum treatment, only the latter was able to clarify the physical reason of the polarization change effect in free space. Namely, the different evolutions of two independent subensembles contained in a photonic mixed state. This result is being presented here for the first time.  
\begin{figure}[htb]
\begin{center}
\includegraphics[height=1.9in]{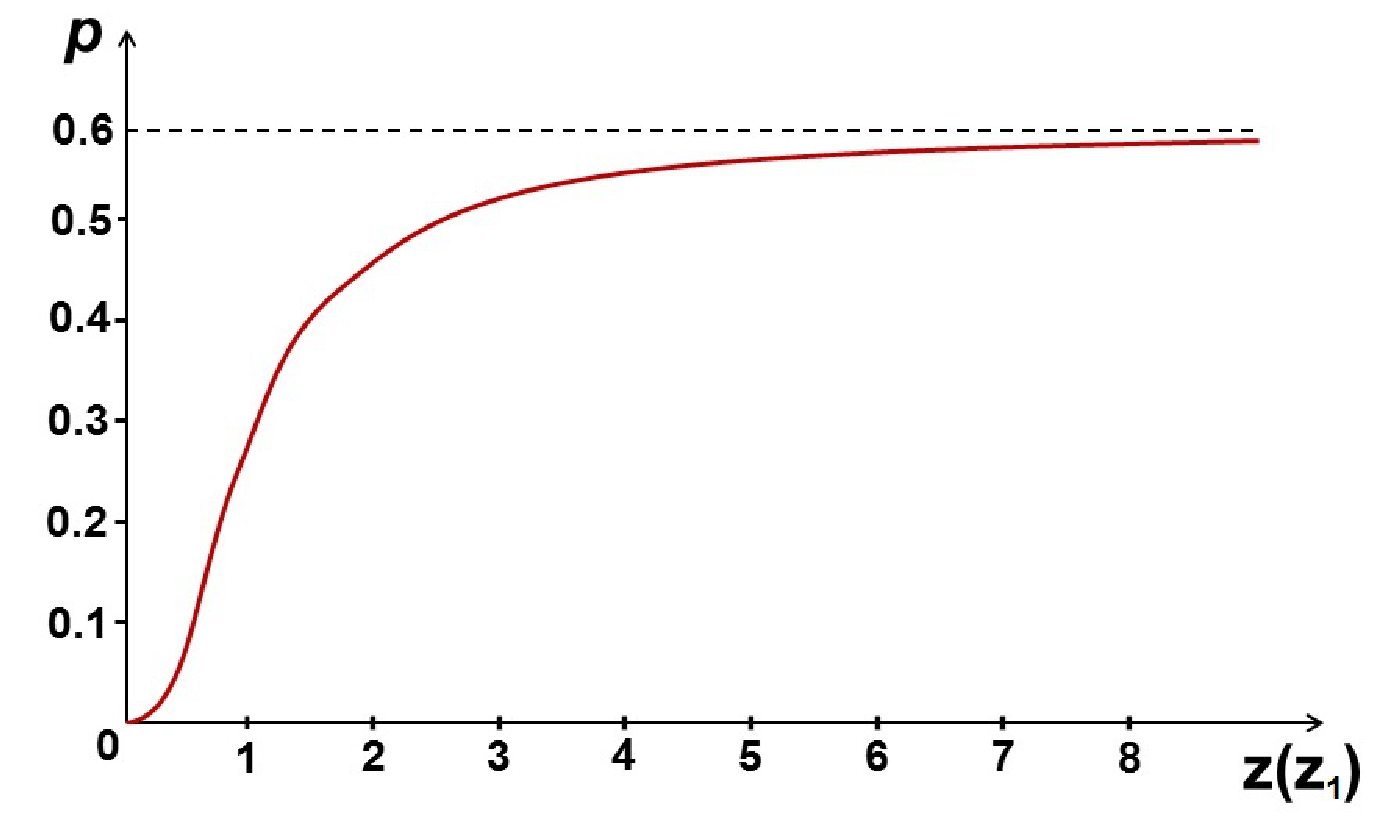}
\end{center}
\label{F1}
\caption{(Color online) Degree of polarization $p$ as a function of the propagation distance $z$ (in unities of $z_{1}$), with $z_{2}=2 z_{1}$, for the ensemble of photons described by Eqs.~\eqref{25} and~\eqref{30}. Observe that $p=0$ for $z=0$ because the fractional probabilities to find each (orthogonal) subensemble is equal, which characterizes unpolarized light. On the other hand, for $z>7z_{1}$ the degree of polarization converges asymptotically to $0.6$. This is because, after the evolution, the fractional probabilities are different, but the ratio between them converges to a fixed value. This is a characteristic of partially polarized light.}
\end{figure}

At this stage, we can point out that the unified theory of polarization and coherence was fundamental in the present description in the sense that, to completely describe the dynamics of the degree of polarization of the photons upon propagation, information about the influence of the dynamics of the phase relation between the two subensembles, which is related the (spatial) coherence properties of the whole ensemble, has to be taken into account. In this regard, the density matrix of Eq.~\eqref{25} is the mathematical entity capable of providing such complete knowledge, in a similar fashion to that expected from the cross-spectral density matrix in the unified classical theory \cite{wolf}.      

\section{Decoherence and depolarization due to environmental interactions}

In this section we analyze the effects of the correlations naturally created between the ensemble of photons and the environment constituents that may take place upon propagation. This effect, so-called environment induced decoherence, has been recognized as the responsible for the emergence of the classical behavior of light and matter from the underlying quantum substrate  \cite{zurek,schloss,schloss2,bert}. In this sense, as we shall see, the decay of the coherence and polarization degrees due to the interaction with the environment appears to be a natural and irreversible process. In mathematical terms, the characteristic trait of decoherence is the decay of the off-diagonal elements of the reduced density matrix of the system obtained from the partial trace of the system-environment density matrix with respect to the environmental states \cite{zurek2,zurek3}.

In order to account for this effect on the photonic states described here, we will describe the decoherence process in the formalism of quantum channels \cite{kraus,nielsen}. One of the advantages of the method, also called {\it operator-sum} formalism, is that the influence of the environment on the photons can be described without specific reference to the interaction Hamiltonian. Thus, we initiate this study by defining the system-environment density matrix under the assumption that they are initially uncorrelated. In this form,
\begin{equation}
\label{31}
\hat{\rho}(0)=\hat{\rho}_{S}(0) \otimes \hat{\rho}_{E}(0),
\end{equation}
where the density matrix of the system $\rho_{S}(0)$ contains information of both coherence and polarization, as in Eq.~\eqref{19}, and the density matrix of the environment can be written in the diagonal decomposition as $\rho_{E}(0) = \sum_{i}w_{i}\ket{E_{i}}\bra{E_{i}}$, with $w_{i}$ as the fractional populations of the environmental states in the basis $\{\ket{E_{i}}\}$. If we consider that the photons and the environment form an isolated system, they necessarily evolve under an unitary operation $\hat{U}(t)$. Then, the evolution of the reduced density matrix of the photons is given by    
\begin{equation}
\label{32}
\hat{\rho}_{S}(t)= Tr_{E} \left\{ \hat{U}(t)  \left[ \hat{\rho}_{S}(0) \otimes \left( \sum_{i}w_{i}\ket{E_{i}}\bra{E_{i}} \right) \right]  \hat{U}^{\dagger}(t)  \right\},
\end{equation}
where $Tr_{E}$ denotes the partial trace over the states of the environment. This equation can be written as  
\begin{equation}
\label{33}
\hat{\rho}_{S}(t)= \sum_{ij} \hat{K}_{ij} \hat{\rho}_{S} \hat{K}^{\dagger}_{ij},
\end{equation}
where $\hat{K}_{ij}$ are the so-called {\it Kraus operators}, which are given by
\begin{equation}
\label{34}
\hat{K}_{ij} = \sqrt{w_{i}} \braket{E_{j}|\hat{U}(t)|E_{i}}.
\end{equation}
It is easy to show that the Kraus operators obey the relation $ \sum_{ij} \hat{K}_{ij} \hat{K}^{\dagger}_{ij} = \mathbb{I}$, where $\mathbb{I}$ is the identity matrix in the Hilbert space of the system $S$.

Now, for us to proceed, it is necessary to describe the characteristics of the environment to be experienced by the photons. As a first example, let us assume that the photons propagate through a region in space composed of many small transparent particles, randomly distributed in space, with an index of refraction different from that of vacuum. This type of environment, which could simulate the lower atmosphere \cite{weich}, as well as impurities in optical fibers \cite{wai}, causes random, uncorrelated phase shifts in the photon quantum state at both points $Q_{0}$ and $Q_{1}$. Each possible single interaction of this type can be described by a unitary transformation that modifies uniquely the state of the environment in the following form:  
\begin{equation}
\label{35}
\ket{H,0}\ket{E_{0}} \rightarrow \sqrt{1-\mathcal{P}} \ket{H,0}\ket{E_{0}} +  \sqrt{\mathcal{P}} \ket{H,0}\ket{E_{1}},  
\end{equation}
\begin{equation}
\label{36}
\ket{H,1}\ket{E_{0}} \rightarrow \sqrt{1-\mathcal{P}} \ket{H,1}\ket{E_{0}} +  \sqrt{\mathcal{P}} \ket{H,1}\ket{E_{2}},  
\end{equation}
\begin{equation}
\label{37}
\ket{V,0}\ket{E_{0}} \rightarrow \sqrt{1-\mathcal{P}} \ket{V,0}\ket{E_{0}} +  \sqrt{\mathcal{P}} \ket{V,0}\ket{E_{1}},
\end{equation}
\begin{equation}
\label{38}
\ket{V,1}\ket{E_{0}} \rightarrow \sqrt{1-\mathcal{P}} \ket{V,1}\ket{E_{0}} +  \sqrt{\mathcal{P}} \ket{V,1}\ket{E_{2}}.  
\end{equation}
The parameter $\mathcal{P}$ in the above equations represent the probability for an interaction between a photon and an environment constituent to occur during a given time interval, $\Delta t$. The states $\ket{E_{0}}$, $\ket{E_{1}}$ and $\ket{E_{2}}$ represent the initial state of the environment, and the states of the environment after the interaction with one photon at $Q_{0}$ and at $Q_{1}$, respectively. Note that, in the present case, the change of the states of the environment does not depend on the polarization of the photon, but only on the localization of the interaction. Physically, we can attribute the change of the environmental state upon interaction to the momentum imparted to the atoms of the environment due to either the scattering of the photon or absorption and reemission of it. Also, note that this kind of interaction is unable to cause a transition in the basis $\{ \ket{0},\ket{1} \}$, i.e., a photon at $Q_{0}$ cannot be sent to $Q_{1}$ due to the environment, and vice-versa.

Before evaluating the Kraus operators of Eq.~\eqref{34}, we call attention to the fact that in the present example $\hat{\rho}_{E} (0)= \sum_{i}w_{i}\ket{E_{i}}\bra{E_{i}} = \ket{E_{0}}\bra{E_{0}}$. Then, from  Eqs.~\eqref{33} and~\eqref{34} we have that the time-dependent density matrix can be simplified to
\begin{equation}
\label{39}
\hat{\rho}_{S}(t)= \sum^{2}_{j=0} \hat{K}_{j} \hat{\rho}_{S} \hat{K}^{\dagger}_{j}
\end{equation}
with
\begin{equation}
\label{40}
\hat{K}_{j} = \braket{E_{j}|\hat{U}(t)|E_{0}}.
\end{equation}
Accordingly, by using Eqs.~\eqref{35} to~\eqref{38} we find that the three possible Kraus operators in the $\{\ket{H,0};\ket{H,1};\ket{V,0};\ket{V,1}\}$ basis are given by            
\begin{eqnarray}
\label{41}
\hat{K}_{0} &=& \sqrt{1-\mathcal{P}}[\ket{H,0}\bra{H,0} + \ket{H,1}\bra{H,1}  \nonumber \\ &+& \ket{V,0}\bra{V,0} + \ket{V,1}\bra{V,1}],  
\end{eqnarray}
\begin{equation}
\label{42}
\hat{K}_{1} = \sqrt{\mathcal{P}} [\ket{H,0}\bra{H,0} + \ket{V,0}\bra{V,0}],  
\end{equation}
\begin{equation}
\label{43}
\hat{K}_{2} = \sqrt{\mathcal{P}} [\ket{H,1}\bra{H,1} + \ket{V,1}\bra{V,1}].  
\end{equation}
Then, by using these results in  Eq.~\eqref{39} we have that the evolution of the density matrix after a time $\Delta t$ is given by
\begin{equation}
\label{44}
\hat{\rho}_{S}=
\begin{pmatrix}
\rho_{11} & (1-\mathcal{P})\rho_{12} & \rho_{13} & (1-\mathcal{P})\rho_{14} \\
(1-\mathcal{P})\rho_{21} & \rho_{22} & (1-\mathcal{P})\rho_{23} & \rho_{24} \\
\rho_{31} & (1-\mathcal{P})\rho_{32} & \rho_{33} & (1-\mathcal{P})\rho_{34} \\
(1-\mathcal{P})\rho_{41} & \rho_{42} & (1-\mathcal{P})\rho_{43} & \rho_{44}
  \end{pmatrix}.
\end{equation}
If this operation is applied $n$ times in succession, the $(1-\mathcal{P})$ terms in the matrix above become $(1-\mathcal{P})^{n}$. Also, if we assume that the interaction probability $\mathcal{P}$ in the time interval $\Delta t$ is of the form $\Gamma \Delta t$, with $\Gamma$ as the probability of an interaction between a photon and an environment constituent per unit time, then, after a time $t = n \Delta t$, we have that $(1-\mathcal{P})^{n} = (1-\Gamma t/n)^{n}$. Thus, for $n \rightarrow \infty$ we obtain $(1-\mathcal{P})^{n} \approx e^{-\Gamma t}$ \cite{preskill}. Therefore, the time evolution of the density matrix can be written as  
\begin{equation}
\label{45}
\hat{\rho}_{S} (t)=
\begin{pmatrix}
\rho_{11} & \rho_{12}e^{-\Gamma t} & \rho_{13} & \rho_{14}e^{-\Gamma t} \\
\rho_{21}e^{-\Gamma t} & \rho_{22} & \rho_{23}e^{-\Gamma t} & \rho_{24} \\
\rho_{31} & \rho_{32}e^{-\Gamma t} & \rho_{33} & \rho_{34}e^{-\Gamma t} \\
\rho_{41}e^{-\Gamma t} & \rho_{42} & \rho_{43}e^{-\Gamma t} & \rho_{44}
  \end{pmatrix}.
\end{equation}
Now, from Eq.~\eqref{10}, we obtain that the degree of coherence of the ensemble of photons under the interaction with the environment decays in the following form:
\begin{equation}
\label{46}
\mu(t) = \frac{(\rho_{12} + \rho_{34})e^{-\Gamma t}}{\sqrt{\rho_{11} + \rho_{33}} \sqrt{\rho_{22} + \rho_{44}}} = \mu(0)e^{-\Gamma t}.
\end{equation}
On the other hand, it is easy to see that the degrees of polarization, $p_{0}$ and $p_{1}$, given by Eqs.~\eqref{16} and~\eqref{17} remain constant. This was expected, since the environment only disturbs the relative phase of the state of the photons with respect to the points $Q_{0}$ and $Q_{1}$.

As a last example, we analyze the decoherence effect on the photons due to an environment whose constituents, besides causing random phase shifts to the photonic states as in the previous case, are also birefringent, i.e., the phase shifts now depend on the polarization state of the photon. In this case, after interaction, the photons modify the environment state in a form that depends both on the localization, $\ket{0}$ and $\ket{1}$, and the polarization state, $\ket{H}$ and $\ket{V}$. Under these assumptions, each possible single interaction can be described by a unitary transformation that changes the initial state of the environment as
\begin{equation}
\label{47}
\ket{H,0}\ket{E_{0}} \rightarrow \sqrt{1-\mathcal{P}} \ket{H,0}\ket{E_{0}} +  \sqrt{\mathcal{P}} \ket{H,0}\ket{E_{1}},  
\end{equation}
\begin{equation}
\label{48}
\ket{H,1}\ket{E_{0}} \rightarrow \sqrt{1-\mathcal{P}} \ket{H,1}\ket{E_{0}} +  \sqrt{\mathcal{P}} \ket{H,1}\ket{E_{2}},  
\end{equation}
\begin{equation}
\label{49}
\ket{V,0}\ket{E_{0}} \rightarrow \sqrt{1-\mathcal{P}} \ket{V,0}\ket{E_{0}} +  \sqrt{\mathcal{P}} \ket{V,0}\ket{E_{3}}  ,
\end{equation}
\begin{equation}
\label{50}
\ket{V,1}\ket{E_{0}} \rightarrow \sqrt{1-\mathcal{P}} \ket{V,1}\ket{E_{0}} +  \sqrt{\mathcal{P}} \ket{V,1}\ket{E_{4}}.  
\end{equation}
Again, $\mathcal{P}$ is the interaction probability. In this form, from Eq.~\eqref{40} we can find the five possible Kraus operators related to this case
\begin{eqnarray}
\label{51}
\hat{K}_{0} &=& \sqrt{1-\mathcal{P}}[\ket{H,0}\bra{H,0} + \ket{H,1}\bra{H,1}  \nonumber \\ &+& \ket{V,0}\bra{V,0} + \ket{V,1}\bra{V,1}],  
\end{eqnarray}
\begin{equation}
\label{52}
\hat{K}_{1} = \sqrt{\mathcal{P}} \ket{H,0}\bra{H,0},  
\end{equation}
\begin{equation}
\label{53}
\hat{K}_{2} = \sqrt{\mathcal{P}} \ket{H,1}\bra{H,1},
\end{equation}
\begin{equation}
\label{54}
\hat{K}_{3} = \sqrt{\mathcal{P}} \ket{V,0}\bra{V,0},
\end{equation}
\begin{equation}
\label{55}
\hat{K}_{4} = \sqrt{\mathcal{P}} \ket{V,1}\bra{V,1}.
\end{equation}
Thus, by substitution of Eqs.~\eqref{52} to~\eqref{55} into the time dependent expression for the density matrix,
\begin{equation}
\label{56}
\hat{\rho}_{S}(t)= \sum^{4}_{j=0} \hat{K}_{j} \hat{\rho}_{S} \hat{K}^{\dagger}_{j},
\end{equation}
we have that
\begin{equation}
\label{57}
\hat{\rho}_{S}=
\begin{pmatrix}
\rho_{11} & (1-\mathcal{P})\rho_{12} &  (1-\mathcal{P})\rho_{13} & (1-\mathcal{P})\rho_{14} \\
(1-\mathcal{P})\rho_{21} &  \rho_{22} & (1-\mathcal{P})\rho_{23} &  (1-\mathcal{P})\rho_{24} \\
 (1-\mathcal{P})\rho_{31} & (1-\mathcal{P})\rho_{32} & \rho_{33} & (1-\mathcal{P})\rho_{34} \\
(1-p)\rho_{41} &  (1-\mathcal{P})\rho_{42} & (1-\mathcal{P})\rho_{43} & \rho_{44}
  \end{pmatrix},
\end{equation}
which, similar to the previous example, if we assume that the interaction probability is linear with time, $\mathcal{P}(\Delta t) = \Gamma \Delta t$, after many interactions the terms $(1-\mathcal{P})$ becomes approximately $e^{-\Gamma t}$, with $t$ as the time elapsed by the interactions. In this case, the evolution of the density matrix is given by
\begin{equation}
\label{58}
\hat{\rho}_{S} (t)=
\begin{pmatrix}
\rho_{11} & \rho_{12}e^{-\Gamma t} & \rho_{13}e^{-\Gamma t} & \rho_{14}e^{-\Gamma t} \\
\rho_{21}e^{-\Gamma t} & \rho_{22} & \rho_{23}e^{-\Gamma t} & \rho_{24}e^{-\Gamma t} \\
\rho_{31}e^{-\Gamma t} & \rho_{32}e^{-\Gamma t} & \rho_{33} & \rho_{34}e^{-\Gamma t} \\
\rho_{41}e^{-\Gamma t} & \rho_{42}e^{-\Gamma t} & \rho_{43}e^{-\Gamma t} & \rho_{44}
 \end{pmatrix}.
\end{equation}
Therefore, given the temporal evolution of the density matrix, with the formalism introduced here we can evaluate the evolution of the degrees of coherence and polarization with time. In this case, one can easily see from equation Eq.~\eqref{10} that the degree of coherence has an exponential decay, $\mu(t) = \mu(0)e^{-\Gamma t}$, similar to the previous case, Eq.~\eqref{46}. Nevertheless, from Eqs.~\eqref{16} and~\eqref{17}, we can verify that the degrees of polarization also decay with time according to
\begin{equation}
\label{59}
p_{0}= \sqrt{1-\frac{4(\rho_{11}\rho_{33}-\rho_{13}\rho_{31}e^{-2\Gamma t})}{(\rho_{11}+\rho_{33})^{2}}}
\end{equation}
and
\begin{equation}
\label{60}
p_{1}= \sqrt{1-\frac{4(\rho_{22}\rho_{44}-\rho_{24}\rho_{42}e^{-2\Gamma t})}{(\rho_{22}+\rho_{44})^{2}}}.
\end{equation}
In this case, in which the light-environment interaction is ruled by Eqs.~\eqref{47} to~\eqref{50}, we have a polarization-dependent decoherence, i.e., contrary to case of the last section, now the polarization properties cause influence on the coherence of the photons. This is where the importance of the unified theory comes into play. It would also be interesting to analyze the case in which an ensemble of photons in a mixed state like that of Eq.~\eqref{22} propagates in the medium described by Eqs.~\eqref{47} to~\eqref{50}, instead of in free space. In such scenario, one has interplay between the coherence and polarization properties of light. Such cross influence could be depicted under the perspective of the present unified framework.

We want to call attention to the fact that we have provided a simplified model for the interactions by assuming ideal (orthonormal) environmental states to illustrate the validity of the present model. In fact, as indicated above, the simplicity of the method lies in accounting for the evolution of the reduced density matrix of the system without specifying the interaction Hamiltonian with the environment. This approach delivers a compact and practical description for the dynamics of open quantum systems \cite{schloss2}. However, it is not difficult to imagine a situation in which the states $\ket{E_{j}}$ obtained after the interactions in Eqs.~\eqref{35} to~\eqref{38}, as well as in Eq.~\eqref{47} to~\eqref{50}, are not orthonormal as we assumed. If we relax this condition, it can be shown that the decay rate $\Gamma$ in the elements of Eqs.~\eqref{45} and~\eqref{58} are not necessarily the same, which would provide more interesting time evolutions for the degrees of coherence and polarization.  

The knowledge about the dynamical properties of a light beam which propagates through a disturbing environment is very important and finds application in many fields, such as optical communications, remote sensing and radar systems \cite{wang}. However, the usual classical description of this problem may be cumbersome, especially when the environment is turbulent \cite{salem2}. We believe that the quantum density matrix approach developed in this section opens a new avenue for investigations of the coherence and polarization properties of light under the action of many types of environments, once we know the quantum state transformations which rule the interactions of the photons with the environment constituents. Furthermore, as well known from decoherence theory, depending on the type of environment interacting with the system, the master equation formalism can also be applied \cite{bru,rivas}. Here, we also have this option since our formalism stand on the density matrix of the system as the fundamental element.  

Finally, we want to point out that, contrary to the classical framework, our quantum description of the problem can depict the coherence and polarization properties of a subensemble of photons which composes, for example, one party of an entangled multipartite system, e.g., one of the constituents emitted from a EPR or a GHZ source \cite{pan}. In this case, our four-dimensional density matrix used to obtain information about coherence and polarization would be the reduced density matrix of the subensemble of photons, which provides all the measurement statistics. As a matter of fact, the quantum density matrix method accounts for entanglement in the multipartite system both in the position of the photons and the polarization degree of freedom.

\section{Conclusion}
 
In conclusion, we have proposed a new unified theory of coherence and polarization based solely on first-principles quantum mechanical arguments. The theory relies on a density matrix written in terms of position and polarization states of an ensemble of photons, from which we derived expressions for the degrees of coherence and polarization of the system. To confirm the validity and efficiency of the model, it was applied to show how the degree of polarization of a mixed ensemble of photons varies on propagation in free space; a problem that, to our knowledge, has been studied only with basis on the classical electromagnetic theory. Furthermore, we successfully used our method to describe the behavior of the coherence and polarization properties of a generic ensemble of photons subjected to interactions with an external environment. In this case, we showed two examples of interacting environments: one causing random phase shifts at two different points perpendicular to the propagation direction, and another causing polarization-dependent phase shifts. In this context, we used the operator-sum representation to unveil the temporal evolution of the system. However, it is important to emphasize that, depending on the type of interaction in which the photonic system is submitted, the master equation formalism can also be used to describe the dynamics. In future works, we intend to use the present study to investigate the action of other types of environment by using master equations.

\begin{acknowledgements}
The author is grateful to A. S. L. Gomes and Cid B. de Ara\'{u}jo for their hospitality at Universidade Federal de Pernambuco, and to the financial support from the Brazilian funding agency CNPq, Grant Number 309292/2016-6.
\end{acknowledgements}



\begin{thebibliography}{7}
\bibitem{note} These fractional populations concern the probabilities to find photons from each subensemble at the points $Q_{0}$ and $Q_{1}$ (at z=0), not the probabilities to find them in the whole volume which contains the ensemble.
\end{thebibliography}
\end{document}